\documentclass[11pt,conference,doublecolumn]{IEEEtran}
\IEEEoverridecommandlockouts
\usepackage[noadjust]{cite}

\ifCLASSINFOpdf
  \usepackage[pdftex]{graphicx}
\else
\fi

\usepackage[cmex10]{amsmath}
\usepackage{amssymb}   
\usepackage{url}

\usepackage[utf8]{inputenc} 
\usepackage[T1]{fontenc}

\usepackage{mleftright}       
\mleftright                   

\usepackage{booktabs}         

\usepackage{caption}
\usepackage{tikz}
\usepackage{xcolor}
\usetikzlibrary{positioning, arrows.meta, shapes.geometric, fit, backgrounds, calc}
\definecolor{wcol}{RGB}{21,96,130}   
\tikzset{
  box/.style ={draw=black, line width=1.2pt, minimum height=0.82cm, font=\large},
  wire/.style={draw=wcol,  line width=0.9pt},
  lbl/.style ={font=\Large, inner sep=1.5pt},
  ar/.style  ={-{Triangle[length=2mm,width=1.9mm]}},
}

\newtheorem{theorem}{Theorem}

\newtheorem{definition}[theorem]{Definition}

\begin{document}
%


\title{
Lossy Compression, Realism, and Coordination
}
%
%
%



\author{%
  \IEEEauthorblockN{Yassine Hamdi\IEEEauthorrefmark{1},~\IEEEmembership{
  Member,~IEEE} and Deniz G\"{u}nd\"{u}z\IEEEauthorrefmark{1},~\IEEEmembership{Fellow,~IEEE}}\\
  \IEEEauthorblockA{\IEEEauthorrefmark{1}Department of Electrical and Electronic Engineering, Imperial College London, UK,}
  \IEEEauthorblockA{\{y.hamdi, d.gunduz\}@imperial.ac.uk}
}


\maketitle

\begin{abstract}
Classical rate-distortion theory characterizes the fundamental limits of lossy compression under fidelity constraints, but minimizing distortion often yields perceptually unsatisfying reconstructions—blurry images, over-smoothed textures, and unnatural artifacts. This has motivated a growing body of work on compression with realism constraints, which require reconstructions to be statistically indistinguishable from natural signals, giving rise to the three-way rate-distortion-perception (RDP) trade-off.
This paper provides an accessible overview of this emerging area and reveals deep connections to another fundamental problem: distributed coordination under rate-limited communication.
Under strong distribution matching 
formulations, both problems
lead to nearly 
identical information-theoretic characterizations, both 
require common randomness (CR) for optimal performance, 
and both rely on similar analytical tools such as the 
soft covering lemma.
Beyond a unifying perspective, we survey recent 
developments in formalizing realism, including batched 
critics and algorithmic realism,
and propose to transfer 
{such paradigms} to coordination --- illustrating how the 
connection continues to generate new problems.

\end{abstract}



%


\section{Introduction}\label{sec:intro}

Despite decades of progress, our storage, communication, and computation resources remain limited relative to our data-acquisition capabilities, making data compression ubiquitous in the digital age.
Lossy compression,
the art of discarding information that is less relevant for a given application,
is essential, from streaming video to training large-scale machine learning models \cite{2023GunduzSurveyContextAndSemanticsAndTaskOriented}. The mathematical foundations of lossy compression were laid by Shannon
(see, e.g., \cite[Chapter~3]{2011BookElGamalKimNetworkInformationTheory}),
whose rate-distortion theory characterizes the minimum number of bits required to represent a source signal within a specified fidelity level. Shannon's framework has guided the development of practical lossy compression systems for over seven decades, yet it leaves a fundamental question unresolved: what exactly should we preserve?

In Shannon's formulation, fidelity is measured by a distortion function that quantifies the point-wise difference between the source and its \textit{reconstruction}; for example, mean squared error (MSE) for images or audio waveforms. Minimizing distortion ensures the reconstruction remains close to the original in a well-defined metric space, and this approach has proven remarkably successful for many applications. However, practitioners have long observed that low distortion does not always translate to high perceptual quality. {Particularly at low bit rates, distortion-optimal compression produces reconstructions that are technically `close' to the source yet appear unnatural: blurry images, over-smoothed textures, or the disappearance of high-frequency detail.}

This tension between fidelity and perceptual quality has motivated a fundamental rethinking of what lossy compression should achieve. Rather than asking only ``how close is the reconstruction to the source?'', we should also ask ``does the reconstruction look realistic?'' {Accordingly, \textit{perception} measures the statistical \textit{realism} or naturalness of the reconstruction, independent of its point-wise correspondence to the source.} A perceptually optimal reconstruction should be indistinguishable from samples drawn from the true data distribution. In image compression, this means textures appear realistic, edges are sharp, and the overall appearance matches what we expect from natural images, even if specific pixel values deviate from the original. {Generative models have achieved remarkable success
in recent years
in synthesizing realistic images, audio, and video,
and
demonstrated
that highly realistic content can be produced; the challenge for compression is to do so while remaining faithful to a specific source signal.}

It has been recognized (e.g., {\cite{2011LiEtAlMainPaperOnDistributionPreservingQuantization,2019BlauMichaeliRethinkingLossyCompressionTheRDPTradeoff}}) that
achieving high perceptual quality (making reconstructions statistically indistinguishable from real data)
generally requires introducing distortion beyond the minimum achievable at a given rate. {This gives rise to a three-way trade-off among rate, distortion, and perception \cite{2019BlauMichaeliRethinkingLossyCompressionTheRDPTradeoff}.
Note that, in the absence of any distortion constraint, rate and perception are not at odds: unlike distortion, perceptual quality is not related to any source sample; hence,
optimal perception~\cite{2019BlauMichaeliRethinkingLossyCompressionTheRDPTradeoff}
can be achieved even at zero rate,
e.g., by drawing a high-quality reconstruction at random from a fixed database.}
In contrast, satisfying a given distortion constraint requires a certain minimum compression rate.
If we additionally impose a
perception constraint,
the compression rate increases.
We will discuss the rate-distortion-perception (RDP) trade-off at length.

But how should high perceptual quality, or ``realism,'' be formalized mathematically? This question lies at the heart of much recent research \cite{2024TheisUniversalCriticsPositionPaper}. We will present various ways of formalizing perceptual quality requirements, with an emphasis on \textit{distribution matching} constraints: these require the distribution of the reconstruction to be close to that of the source, according to some measure of similarity between distributions. When the two distributions are nearly identical, a large collection of reconstructions
would be statistically indistinguishable from genuine source samples. Such distribution matching has deep intuitive appeal: it captures the idea that reconstructions should `look like the real thing' as an ensemble,
exhibiting the same statistical properties as natural signals.

Indeed, distribution matching constraints are ubiquitous in another area of network information theory: \textit{distributed coordination}.
{Consider two autonomous drones conducting a search operation, each making its own flight-path decisions.} They must coordinate their movements using limited communication, e.g., to maximize coverage. Due to bandwidth or latency constraints, one drone cannot simply transmit its complete flight plan to the other; instead, it must compress relevant information into a short message. The receiving drone must then choose an action (its own flight path) that is appropriately coordinated with the first drone's movement. What does `appropriately coordinated' mean? One natural formalization is that the joint distribution of both drones' actions should match some target distribution representing desirable coordinated behavior.
The target joint distribution $q_{X,Y}$ is chosen by a system 
designer to encode the desired coordination objective: it assigns 
high probability to pairs of flight paths providing complementary 
coverage and low probability to collision-prone configurations.
The communication constraint makes it generally impossible to 
achieve this target exactly at finite rate, giving rise to a 
rate-coordination trade-off.

This is precisely a distribution matching constraint, now applied to the joint behavior of multiple agents rather than to reconstructions of a source signal. A common mathematical formulation, known as \textit{channel synthesis} or \textit{strong coordination}~\cite{Cuff2010CoordinationCapacity,2013PaulCuffDistributedChannelSynthesis}, asks: what is the minimum communication rate required for the decoder (second agent) to produce outputs such that the joint input-output distribution matches a specified target? {This problem seems far removed from perceptual image compression.} Yet, as we shall demonstrate, the two problems share deep structural similarities when the latter is formulated with strong distribution matching requirements:\\
\textit{(i) Near-identical achievable regions:} the 
information-theoretic characterizations of the optimal 
trade-offs differ only in minor ways.\\
\textit{(ii) The necessity of randomization:} both problems 
require randomized encoders and decoders, whereas 
deterministic schemes suffice for distortion-only 
compression.\\
\textit{(iii) The critical role of common randomness (CR):} 
both problems require the encoder and decoder to share CR 
whose entropy may exceed the communication rate.\\
\textit{(iv) Common analytical tools:} proofs of both results rely on the soft 
covering lemma and employ similar coding schemes, notably 
the likelihood encoder.

These parallels suggest that perceptual compression and distributed coordination are, in a precise mathematical sense, two manifestations of the same underlying phenomenon: the challenge of inducing a target probability distribution
through rate-limited communication.
More importantly, these connections have proven 
productive.
Historically, the first characterization of 
the role of CR in compression with realism 
constraints~\cite{Sep2015RDPLimitedCommonRandomnessSaldi} 
was obtained using a proof structure nearly identical to 
the one for channel 
synthesis~\cite{2013PaulCuffDistributedChannelSynthesis}. 
In the presence of side information, the connection 
suggested the definition of a joint realism 
constraint~\cite{OurJournalVersionRDPWithSideInfo}, and 
the proof technique used for channel synthesis was once again 
successful, albeit requiring a soft covering lemma 
tailored to side information. In the present paper, we 
propose to transfer {other paradigms for formalizing the RDP trade-off, namely batched critics and algorithmic realism, to coordination. This leads} to new open problems 
(Table~\ref{tab:constraint_taxonomy}).

\begin{figure*}[t!]
\centering
\begin{tikzpicture}[
    block/.style={
        rectangle, 
        draw, 
        thick,
        minimum width=1.6cm, 
        minimum height=0.8cm,
        align=center,
        fill=white,
        font=\footnotesize
    },
    smallblock/.style={
        rectangle, 
        draw, 
        thick,
        minimum width=1.2cm, 
        minimum height=0.6cm,
        align=center,
        fill=white,
        font=\footnotesize
    },
    arrow/.style={
        ->,
        >=Stealth,
        semithick
    },
    dashedarrow/.style={
        ->,
        >=Stealth,
        semithick,
        dashed
    },
    constraint/.style={
        rectangle,
        draw,
        rounded corners,
        semithick,
        fill=gray!10,
        align=center,
        font=\scriptsize
    },
    region/.style={
        rectangle,
        draw,
        rounded corners,
        semithick,
        fill=gray!5,
        align=center,
        font=\scriptsize
    },
    title/.style={
        font=\small\bfseries
    },
    parallel/.style={
        draw=black!70,
        semithick,
        <->,
        >=Stealth
    },
    paralleltext/.style={
        font=\scriptsize\itshape,
        align=center
    },
    scale=0.85,
    transform shape
]


\node[title] at (0, 3.8) {Rate-Distortion-Perception (RDP)};
\node[font=\scriptsize\itshape] at (0, 3.4) {Lossy Compression with Realism};

\node[smallblock] (source1) at (-3.0, 1.5) {Source};

\node[block, fill=orange!15] (enc1) at (-0.8, 1.5) {Encoder\\[-1pt]$F^{(n)}$};

\node[block, fill=cyan!15] (dec1) at (2, 1.5) {Decoder\\[-1pt]$G^{(n)}$};

\node[smallblock] (out1) at (4.2, 1.5) {Reconstr.};

\node[smallblock, fill=yellow!20] (cr1) at (0.6, 2.8) {\scriptsize CR $J$};

\draw[arrow] (source1) -- node[above, font=\scriptsize] {$X_{1:n}$} (enc1);
\draw[arrow] (enc1) -- node[above, font=\scriptsize] {$M$} node[below, font=\tiny] {$nR$ bits} (dec1);
\draw[arrow] (dec1) -- node[above, font=\scriptsize] {$Y_{1:n}$} (out1);
\draw[dashedarrow] (cr1.south west) -- (enc1.north);
\draw[dashedarrow] (cr1.south east) -- (dec1.north);
\node[font=\tiny, right=0.1cm of cr1] {$nR_c$ bits};

\node[constraint, below=0.6cm of dec1, text width=4.2cm, anchor=north] (const1) {
    \textbf{Constraints:}\\[1pt]
    Distortion: $\mathbb{E}[d(X, Y)] \leq \lambda_1$\\[1pt]
    \fbox{Realism: $p_Y = p_X$}
};

\node[region, below=0.3cm of const1, text width=4.2cm, anchor=north] (reg1) {
    \textbf{Achievable Region:}\\[1pt]
    $R \geq I(X; V)$\\[1pt]
    \fbox{$R + R_c \geq I(Y; V)$}\\[1pt]
    $X - V - Y$
};


\node[title] at (9.5, 3.8) {Channel Synthesis};
\node[font=\scriptsize\itshape] at (9.5, 3.4) {Distributed Coordination};

\node[smallblock] (source2) at (6.5, 1.5) {Input};

\node[block, fill=orange!15] (enc2) at (8.7, 1.5) {Encoder\\[-1pt]$F^{(n)}$};

\node[block, fill=cyan!15] (dec2) at (11.5, 1.5) {Decoder\\[-1pt]$G^{(n)}$};

\node[smallblock] (out2) at (13.7, 1.5) {Output};

\node[smallblock, fill=yellow!20] (cr2) at (10.1, 2.8) {\scriptsize CR $J$};

\draw[arrow] (source2) -- node[above, font=\scriptsize] {$X_{1:n}$} (enc2);
\draw[arrow] (enc2) -- node[above, font=\scriptsize] {$M$} node[below, font=\tiny] {$nR$ bits} (dec2);
\draw[arrow] (dec2) -- node[above, font=\scriptsize] {$Y_{1:n}$} (out2);
\draw[dashedarrow] (cr2.south west) -- (enc2.north);
\draw[dashedarrow] (cr2.south east) -- (dec2.north);
\node[font=\tiny, right=0.1cm of cr2] {$nR_c$ bits};

\node[constraint, below=0.6cm of dec2, text width=4.2cm, anchor=north] (const2) {
    \textbf{Constraint:}\\[1pt]
    ~\\[1pt]
    \fbox{Coordination: $p_{X,Y} = q_{X,Y}$}
};

\node[region, below=0.3cm of const2, text width=4.2cm, anchor=north] (reg2) {
    \textbf{Achievable Region:}\\[1pt]
    $R \geq I(X; V)$\\[1pt]
    \fbox{$R + R_c \geq I(X,Y; V)$}\\[1pt]
    $X - V - Y$
};


\draw[thick, gray!50, dashed] (5.2, 4.2) -- (5.2, -3.2);

\draw[parallel] ($(const1.east)+(0.15,0)$) -- ($(const2.west)-(0.15,0)$);

\draw[parallel] ($(reg1.east)+(0.15,0)$) -- ($(reg2.west)-(0.15,0)$);

\node[paralleltext, fill=white, inner sep=2pt] at (6.2, -0.6) {$Y \leftrightarrow (X,Y)$};

\end{tikzpicture}

\caption{The structural parallel between RDP with strong realism constraints (left) and channel synthesis for distributed coordination (right)  --- formalized in Section \ref{sec:unified_problem_formulation} and described in detail in Section \ref{sec:distribution_matching_everything}. Both problems share identical system architectures with encoder, decoder, and CR. The key difference lies in the distribution matching constraint: RDP requires the marginal distribution $p_Y$ to match $p_X$, while channel synthesis requires the joint distribution $p_{X,Y}$ to match a target $q_{X,Y}$. This single change---replacing $Y$ with $(X,Y)$---is the only substantive difference between the achievable regions, revealing that these seemingly disparate problems are fundamentally the same.}
\label{fig:unified_rdp_coordination}
\end{figure*}
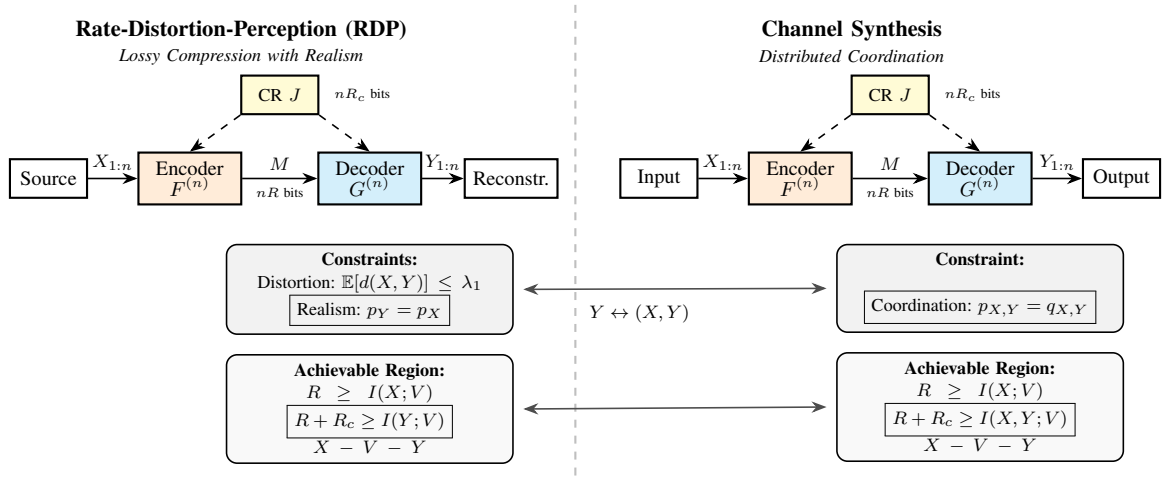

Beyond distribution matching, recent work has explored alternative formalizations of realism that may better align with human perception and practical constraints.
As mentioned previously,
distribution matching constraints essentially require that a large ensemble of reconstructions
be statistically similar to the source. What if, instead, one considers a realism requirement involving a small ensemble of reconstructions, or a single reconstruction? In general, \textit{batched critics}
evaluate realism by inspecting multiple reconstructions simultaneously, interpolating between single-sample evaluation and full distributional constraints. Another paradigm, namely \textit{algorithmic realism}, draws on concepts from algorithmic information theory to define realism through {a large family of} computationally bounded tests, {bridging certain gaps between distribution matching theory and practice~\cite{2025OurISITRDPAlgorithmicRealism}.}
Developments on the RDP side
naturally raise the question of whether analogous frameworks exist for coordination, a question we discuss in Section~\ref{sec:algorithmic_realism}, illustrating the broader thesis that the RDP-coordination connection is a continual source of new problems and insights.

\textbf{Organization of the paper:} {This paper aims to provide an accessible overview of lossy compression with realism constraints}
and distributed coordination theory, organized around their deep structural connections.
In Section \ref{sec:unified_problem_formulation}, we present a unified mathematical formalism that encompasses rate-distortion theory, the RDP trade-off, and distributed coordination as special cases,
and we introduce the taxonomy of constraints that underpins 
the parallel (Table~\ref{tab:constraint_taxonomy}).
In Section \ref{sec:distribution_matching_everything}, we
present
some of
the known achievable regions for
both 
problem classes side by side,
{along with the quantities that govern the role of common randomness,}
across
the point-to-point setting,
remote source compression, and compression with side information.
In Section \ref{sec:algorithmic_realism}, we discuss batched critics and algorithmic realism
--- recent advances on the RDP side --- and 
argue that they exemplify the productive transfer of ideas across 
the parallel by suggesting open problems for coordination.
Throughout, we emphasize intuition and connections over technical proofs, aiming to provide a reference for researchers at the intersection of lossy compression, distributed coordination, and generative modeling.

\section{A Unified Problem formulation}\label{sec:unified_problem_formulation}

This section develops a framework broad enough to capture Shannon's rate-distortion theory, the RDP trade-off, and distributed coordination as special cases; casting them in a common formalism is what lets us expose their structural similarities in Section~\ref{sec:distribution_matching_everything}. Readers primarily interested in the conceptual parallels may skim the technical details on first reading and return to them as needed.

\medskip
\noindent\textbf{On the role of the i.i.d.\ model.}
Throughout this paper, we model the source as a sequence of
independent and identically distributed (i.i.d.)
symbols. Natural signals such as images are not i.i.d., 
and practical compression operates on individual signals rather 
than asymptotically long sequences. The i.i.d.\ block coding 
model is an idealization whose value lies in its tractability: 
it enables precise characterizations of fundamental trade-offs 
and reveals structural properties, such as the role of 
randomization, the necessity of common randomness, and the 
deep parallels between realism and coordination, which would have been difficult to discover in more complex models. It has also proven a productive starting point throughout information theory~{\cite{2011BookElGamalKimNetworkInformationTheory}}: from the source coding theorem, through the rate-distortion function, to channel capacity, insights first established in this idealized regime have informed both the theory and practice of compression and communication.

\subsection{The Basic Setup}

For clarity of exposition, we consider a point-to-point communication system with two terminals: an \textit{encoder} that observes a source and transmits a compressed message, and a \textit{decoder} that receives the message and produces an output. The setup is depicted in Figure~\ref{fig:unified_rdp_coordination}.

The source is a sequence $X_{1:n}= (X_1, X_2, \ldots, X_n)$ of $n$ i.i.d. symbols, each drawn from alphabet $\mathcal{X}$ according to distribution $p_X.$ We write this as $X_{1:n} \sim p_X^{\otimes n}$.
The encoder compresses this source sequence into a message $M$ consisting of approximately $nR$ bits, where $R$ is called the \emph{compression rate} (measured in bits per source symbol). The decoder receives $M$ and produces an output sequence $Y_{1:n}$ on alphabet $\mathcal{Y}$. In rate-distortion problems, $Y_{1:n}$ is called the \textit{reconstruction} sequence and we typically have $\mathcal{Y}=\mathcal{X}$; whereas in coordination problems, $Y_{1:n}$ represents the \emph{actions} of a second agent.

A key feature of our formulation, one that distinguishes it from classical rate-distortion theory, is that both the encoder and decoder may be \emph{randomized}. That is, rather than being deterministic functions, they may be stochastic mappings that use local random number generators. Moreover, they may have access to  \emph{common randomness} (CR): a random variable $J$ that is available to both terminals before communication begins.
We assume the CR is genuinely random, not the output of a pseudo-random generator. One can think of a pseudo-random number generator but with a truly \textit{random} (i.e., unpredictable) state --- or with a truly random seed: {that is, a fixed sequence $\{J^{(k)}\}_{1 \leq k \leq \lfloor 2^{nR_c} \rfloor}$ such that for every new source sample $X_{1:n}$ processed by the encoder, both terminals can agree on a new random index $K$}
independently of the content of $X_{1:n}$ and $M$\footnote{For instance, the decoder may infer the value of $K$ from the knowledge that a new message $M$ has been sent by the encoder and from an external source of CR, but should not need to read the content of $M$ to determine $K$ or $J^{(K)}.$}, where $K$ follows a uniform distribution.
As we shall see, this CR plays a crucial role in problems with stringent distribution matching constraints. This is in stark contrast to classical rate-distortion theory, where deterministic schemes suffice.

\begin{definition}\label{def:n_letter_code}
Given rates $R,R_c \geq 0,$ and blocklength $n\in\mathbb{N},$
an $(n, R, R_c)$ \textit{code} consists of
an \emph{encoder} $F^{(n)}$:
a conditional distribution 
$F^{(n)}_{M|X_{1:n}, J}$ that maps the source sequence
$X_{1:n}$
and CR $J$ to a message
$M$ in the set

\noindent $[2^{nR}]$\footnote{For any positive real number $a$, 
we denote by $[a]$ the set of integers less than or equal 
to $\lfloor a \rfloor$, i.e., 
$[a] = \{1, 2, \ldots, \lfloor a \rfloor\}$.}, 
and a \emph{decoder} $G^{(n)}$: a conditional 
distribution $G^{(n)}_{Y_{1:n}|M, J}$ that maps the 
message and CR to an output sequence 
$Y_{1:n} \in \mathcal{Y}^n$.
The \emph{induced distribution} of the code is the joint distribution over all variables:
\begin{IEEEeqnarray}{c}
P^{(n)}
\triangleq
p_{X}^{\otimes n} \cdot p^{\mathcal{U}}_{[2^{nR_c}]} \cdot 
F^{(n)}_{M|X_{1:n}, J} \cdot G^{(n)}_{Y_{1:n}|M,J},
\IEEEeqnarraynumspace
\end{IEEEeqnarray}
where $p^{\mathcal{U}}_{[2^{nR_c}]}$ denotes the uniform distribution over $[2^{nR_c}]$ (the distribution of the CR $J$).

A code is said to be \textit{deterministic} if both encoder and decoder are deterministic
and
do not depend on $J$:
\begin{IEEEeqnarray}{c}
\forall x_{1:n}, \exists\, m \text{ such that } F^{(n)}_{M|X_{1:n}, J}(m \,|\, x_{1:n}, j) = 1 \ \forall j,
\nonumber
\\
\forall m, \exists\, y_{1:n} \text{ such that } G^{(n)}_{Y_{1:n}|M, J}(y_{1:n} \,|\, m, j) = 1 \ \forall j.
\nonumber
\end{IEEEeqnarray}

\end{definition}

The CR $J$ takes $\lfloor 2^{nR_c} \rfloor$ possible values, so $R_c$ represents the rate of shared randomness in bits per source symbol. When $R_c = 0$ and the code is deterministic, we recover the classical setup of Shannon's rate-distortion theory.

\subsection{Enter: Realism Constraints}

{While Shannon's rate-distortion theory has driven the design of countless compression systems, it has a basic limitation: distortion measures are imperfect proxies for human perception.} This mismatch becomes apparent in several ways.

Consider an i.i.d.\ Gaussian source compressed under MSE 
distortion. The optimal reconstruction has lower variance 
than the source. For stationary Gaussian sources with non-flat power spectrum, the reverse waterfilling procedure \cite{2011BookElGamalKimNetworkInformationTheory} nulls the spectrum at high frequencies. This is 
why JPEG images appear blurry at low bit rates.

More broadly, all distortion measures used in multimedia compression are stand-ins for what we truly care about: how the reconstruction will be perceived by the end user. A reconstruction might achieve low MSE while appearing unnatural---overly smooth, lacking texture, or exhibiting artifacts that the human visual system immediately detects. This has motivated the development of perceptual distortion measures such as SSIM~\cite{2003SSIM} and LPIPS~\cite{2018LPIPS}, which attempt to capture aspects of human perception within the classical rate-distortion framework.
More recently, the Wasserstein distortion 
function~\cite{
2024ISITYangQiuAaronWagnerWassersteinDistortionFirstPublicationOfItsThoereticalProperties} 
has been proposed.
{Inspired by the human visual system, it compares distributions of local features in a latent space.}
It has contributed to a 
state-of-the-art neural compression 
codec~{\cite{2025GoogleWassersteinDistortionToImproveOverfittedCompressor}}.

An alternative approach is to constrain the \emph{statistical properties} of reconstructions directly, rather than designing ever more sophisticated distortion functions. {Natural signals exhibit characteristic statistical regularities.} A realistic reconstruction should possess these regularities, independent of its point-wise correspondence to any particular source sample. This idea of \emph{distribution matching} has a long history, appearing in early work on distribution-preserving quantization~\cite{
2011LiEtAlMainPaperOnDistributionPreservingQuantization}.

The recent success of deep generative models has brought 
renewed interest in this approach. Neural codecs can now
outperform traditional algorithms under various quality 
metrics~\cite{2022PoELIC}, and at very low bit rates, 
generative models such as 
GANs~{\cite{2019AgustssonMentzerGANforExtremeCompression}} or 
diffusion models~{\cite{2022TheisGaussianDiffusion}} can 
synthesize plausible image content that is semantically 
similar to the source but may differ in fine details. 
These developments have revealed a fundamental tension: 
high perceptual quality often requires accepting higher 
distortion, as formalized in the RDP 
trade-off~{\cite{2019BlauMichaeliRethinkingLossyCompressionTheRDPTradeoff}}.

\subsection{
Problem formulation
}\label{subsec:constraints_toolbox}

The main objective of the remainder of Section \ref{sec:unified_problem_formulation} is
to
exhibit the substantial similarity between problems with realism (perception) and coordination constraints in terms of formulation.
We defer the presentation of the similarity in terms of information-theoretic results to Section \ref{sec:distribution_matching_everything}.
The reader
may focus on Sections \ref{subsubsec:distortion_constraints}, \ref{subsubsec:strong_realism_constraints},
\ref{subsubsec:strong_coordination_constraints}, \ref{subsec:unifying_definition_of_achievability}, and \ref{subsec:clarifying_the_abstract_formulation} on first reading.

We organize the problem formulation into three parts: \emph{distortion},
\emph{realism},
and \emph{coordination}.
The calligraphic letter $\mathcal{P}$ stands for (mathematical) \textit{property}.

\subsubsection{Distortion Constraints}\label{subsubsec:distortion_constraints}
Shannon's rate-distortion theory considers the average distortion between the source and reconstruction:
\begin{IEEEeqnarray}{c}
\mathcal{P}_{\mathrm{dist}}(n, P^{(n)}, \lambda): \mathbb{E}_{P^{(n)}}\bigl[d(X_{1:n}, Y_{1:n})\bigr] \leq \lambda,
\label{eq:def_distortion_n_letter}
\IEEEeqnarraynumspace
\end{IEEEeqnarray}
where $d$ measures the discrepancy between the source sequence and its reconstruction. The most common choice is an \emph{additive} (or single-letter) distortion:
\begin{IEEEeqnarray}{c}
d(X_{1:n}, Y_{1:n}) = \frac{1}{n} \sum_{t=1}^{n} d(X_t, Y_t),
\label{eq:def_additive_distortion_function}
\end{IEEEeqnarray}
where $d: \mathcal{X} \times \mathcal{X} \to [0, \infty)$ is a single-letter distortion measure such as squared error $d(x, y) = (x - y)^2$ or Hamming distance $d(x, y) = \mathbf{1}\{x \neq y\}$.

A central result in rate-distortion theory is that the optimal trade-off between rate $R$ and distortion $\lambda$ can be achieved by \emph{deterministic} codes; that is, CR provides no benefit. This can change dramatically when we add certain realism constraints.

\subsubsection{Realism Constraints}

The most studied realism constraints require the reconstruction to be statistically similar to natural signals, independent of its correspondence to any particular source sample. {The idea is \emph{distribution matching}: the distribution of reconstructions should resemble that of the sources.}

\paragraph{Strong Realism Constraints}\label{subsubsec:strong_realism_constraints}

The most stringent form requires the joint distribution of the entire reconstruction sequence to match that of the source \cite{2011LiEtAlMainPaperOnDistributionPreservingQuantization,Sep2015RDPLimitedCommonRandomnessSaldi,Dec2022WeakAndStrongPerceptionConstraintsAndRandomness}:
\begin{IEEEeqnarray}{c}
\mathcal{P}_{\mathrm{strong}}(n, P^{(n)}, \lambda): \mathcal{D}_n\bigl(P^{(n)}_{Y_{1:n}}, p_X^{\otimes n}\bigr) \leq \lambda,
\label{eq:def_strong_realism}
\end{IEEEeqnarray}
where $\mathcal{D}_n$ is a measure of discrepancy between distributions over $\mathcal{Y}^n = \mathcal{X}^n$, such as total variation distance (TVD). When $\lambda = 0$, this requires $P^{(n)}_{Y_{1:n}} = p_X^{\otimes n}$: the reconstruction sequence must be statistically indistinguishable from an i.i.d.\ source sequence.

\paragraph{Per-Symbol Realism Constraints}
A weaker requirement constrains only the marginal distributions of individual symbols \cite{TheisWagner2021VariableRateRDP,Dec2022WeakAndStrongPerceptionConstraintsAndRandomness}:

\begin{IEEEeqnarray}{c}
\mathcal{P}_{\mathrm{per\text{-}symbol}}(n, P^{(n)}, \lambda):
\max_{1\le t \le n}
\mathcal{D}\bigl(P^{(n)}_{Y_t}, p_X\bigr) \leq \lambda,
\label{eq:def_per_symbol_realism}
\IEEEeqnarraynumspace
\end{IEEEeqnarray}
where $\mathcal{D}$ measures discrepancy between distributions over $\mathcal{X}$. One could also consider the variant with an average over $t$ instead of a maximum. This type of constraint ensures each reconstruction symbol has approximately the right distribution, but does not constrain dependencies between symbols. In particular, it is invariant with respect to the order of symbols.

\paragraph{Empirical Realism Constraints}
In Shannon's rate-distortion scheme, the reconstruction sequence $Y_{1:n}$ is, with high probability, \emph{typical}; that is,
its empirical statistics converge to population statistics as $n$ grows:
\begin{IEEEeqnarray}{c}
\mathcal{D}(\hat{P}_{Y_{1:n}}, q) \overset{\mathbb{P}}{\longrightarrow} 0,
\label{eq:typicality_convergence}
\end{IEEEeqnarray}
for some distribution $q$ on $\mathcal{Y},$
where
the \emph{empirical distribution}
$\hat{P}_{y_{1:n}}$
of a sequence $y_{1:n}$
is
the frequency distribution (on $\mathcal{Y}$) of its
symbols,
and
$\overset{\mathbb{P}}{\longrightarrow}$ means that for any $\varepsilon > 0$, the probability that $\mathcal{D}(\hat{P}_{Y_{1:n}}, q) > \varepsilon$ vanishes as $n \to \infty$. This motivates constraints that require reconstructions to have the ``right'' empirical statistics.
An \textit{empirical realism} constraint requires
that $\hat{P}_{Y_{1:n}}$
be close to $p_X$ {\cite{Sep2015RDPLimitedCommonRandomnessSaldi,Dec2022WeakAndStrongPerceptionConstraintsAndRandomness}}. It can take the form \eqref{eq:typicality_convergence} with $q=p_X,$ or
involve a requirement on every individual realization:
\begin{IEEEeqnarray}{c}
\mathcal{P}_{\mathrm{empr}}(n, P^{(n)}, \lambda): P^{(n)}\bigl(\mathcal{D}(\hat{P}_{Y_{1:n}}, p_X) \leq \lambda\bigr) = 1.
\label{eq:def_empirical_realism}
\IEEEeqnarraynumspace
\end{IEEEeqnarray}
Like the per-symbol constraint, it is invariant to the order of symbols.

\subsubsection{Coordination Constraints}

In coordination problems, the goal is not to reconstruct the source, but to generate an output $Y_{1:n}$ that is appropriately correlated with the input $X_{1:n}$ according to some target distribution $q_{X,Y}$. This models scenarios where two agents must coordinate their actions through limited communication.
The problem of generating samples from a chosen target joint distribution in a distributed manner has long been of interest 
\cite{1973GacsKornerCommonInformation, 1975WynerCommonInformationAndSoftCoveringLemma, Cuff2010CoordinationCapacity}.

\paragraph{Strong Coordination Constraints}\label{subsubsec:strong_coordination_constraints}
The most stringent form requires the joint distribution of $(X_{1:n}, Y_{1:n})$ to match the i.i.d. target $q_{X,Y}^{\otimes n}$:
\begin{IEEEeqnarray}{c}
\mathcal{P}_{\mathrm{strong\text{-}coord}}(n, P^{(n)}, \lambda): \mathcal{D}_n\bigl(P^{(n)}_{X_{1:n}, Y_{1:n}}, q_{X,Y}^{\otimes n}\bigr) \leq \lambda.
\label{eq:def_strong_coordination}
\IEEEeqnarraynumspace
\end{IEEEeqnarray}
When $\lambda = 0$, the encoder-decoder pair effectively implements a memoryless channel $q_{Y|X}$ from input to output, despite the rate constraint. This problem is known as \emph{channel synthesis}.

\paragraph{Empirical Coordination Constraints}
The weaker empirical version requires the joint empirical distribution of $(X_{1:n}, Y_{1:n})$ to converge to the target:
\begin{IEEEeqnarray}{c}
\mathcal{P}_{\mathrm{emp\text{-}coord}}\bigl(\{P^{(n)}\}\bigr): \mathcal{D}\bigl(\hat{P}_{X_{1:n}, Y_{1:n}}, q_{X,Y}\bigr) \overset{\mathbb{P}}{\longrightarrow} 0,
\label{eq:def_empirical_coordination}
\IEEEeqnarraynumspace
\end{IEEEeqnarray}
where $\hat{P}_{X_{1:n}, Y_{1:n}}$ is the joint empirical distribution (frequency of pairs), and $\overset{\mathbb{P}}{\longrightarrow}$ denotes convergence in probability under $P^{(n)}$.

\medskip
\noindent\textbf{Remark.} Comparing~\eqref{eq:def_strong_realism} and~\eqref{eq:def_strong_coordination} reveals the structural similarity between realism and coordination constraints. In both cases, we require a distribution induced by the code to match a target distribution. The difference lies in \emph{which} distribution: 
\begin{itemize}
\setlength{\itemsep}{0.3em}
    \item Realism: marginal $P^{(n)}_{Y_{1:n}}$ should match $p_X^{\otimes n}$;
    \item Coordination: joint $P^{(n)}_{X_{1:n}, Y_{1:n}}$ should match $q_{X,Y}^{\otimes n}$.
\end{itemize}
{The observation that $Y$ is replaced by $(X, Y)$ is central to the parallels we explore in Section~\ref{sec:distribution_matching_everything}.} {Figure~\ref{fig:unified_rdp_coordination} illustrates this correspondence.}

\subsubsection{Realization-Based Constraints and Batched Critics}

Intuitively, the distortion constraint \eqref{eq:def_distortion_n_letter} and the strong realism constraint \eqref{eq:def_strong_realism} differ fundamentally in how they depend on the induced distribution $P^{(n)}.$
Equation~\eqref{eq:def_distortion_n_letter} is of the form
\begin{IEEEeqnarray}{c}
\mathbb{E}[\phi^{(n)}(\text{samples from }P^{(n)})] \leq \lambda,
\end{IEEEeqnarray}
where $\phi^{(n)}$ is a fixed function (no dependence on $P^{(n)}$) and $\lambda$ is a constant.
{We call any such constraint a \textit{realization-based constraint} and $\phi^{(n)}$ a \textit{realization-based metric}; we also allow inequalities} of the form $\limsup_n \ \mathbb{E}[\phi^{(n)}(\text{samples from }P^{(n)})] < \lambda,$ where $\lambda$ may be $\infty.$
Note that the expectation becomes a probability if $\phi^{(n)}$ is an indicator function.
Examples include constraints \eqref{eq:def_distortion_n_letter}, \eqref{eq:def_empirical_realism}, and \eqref{eq:def_empirical_coordination}
(the 
latter through the equivalence, for bounded $\mathcal{D}$, 
between convergence in probability and convergence of 
expectations).
All other constraints are called \textit{distributional constraints}.
For most divergences $\mathcal{D},\mathcal{D}_n,$ such as the TVD,
the constraints
\eqref{eq:def_strong_realism},
\eqref{eq:def_per_symbol_realism}, and \eqref{eq:def_strong_coordination}
are not realization-based.

A realization-based metric may jointly inspect multiple reconstructions:
\begin{IEEEeqnarray}{c}
\mathbb{E}
[\delta(Y^{(1)}_{1:n},\ldots,Y^{(B_n)}_{1:n})],
\label{eq:batched_critic}
\end{IEEEeqnarray}
where $\{Y^{(k)}_{1:n}\}_{k \in [B_n]}$ is a batch of $B_n$ i.i.d.\ samples from $P^{(n)}_{Y_{1:n}}$, and $\delta$ is a real-valued function. Naturally, any such function is called a \emph{batched critic}~\cite{2024TheisUniversalCriticsPositionPaper}.
Note that the coding scheme itself does not involve any batch; only the realism metric does.
The key property of this class of metrics is that
{the batch size $B_n$ effectively interpolates between two fundamentally different constraints: realization-based metrics with a single sample ($B_n = 1$) and full distributional constraints.}
Indeed,
for certain choices of $\delta,$ a realization-based constraint involving \eqref{eq:batched_critic} is equivalent to a distributional constraint in the limit $B_n \to \infty.$ {Heuristically, inspecting a large batch of reconstructions jointly amounts to examining their full distribution $P^{(n)}_{Y_{1:n}}.$}
In Section~\ref{sec:algorithmic_realism}, we present an application of this paradigm to the RDP problem, for which the role of randomness varies dramatically with $B_n.$ 

The batched critics paradigm encompasses a vast set of realization-based metrics: in an RDP setup, the samples from $P^{(n)}$ are reconstructions, but one could equally consider pairs $(X_{1:n},Y_{1:n}).$
The structural parallel 
between realism and coordination documented in 
Section~\ref{sec:distribution_matching_everything} suggests 
that similar ideas can be translated to coordination, filling 
the open entry in Table~\ref{tab:constraint_taxonomy}. 
We develop this direction in 
Section~\ref{sec:algorithmic_realism}, after introducing 
the necessary formalism.

\begin{table}[t]
\centering
\renewcommand{\arraystretch}{1.4}
\setlength{\tabcolsep}{4pt}
\begin{tabular}{@{}l c c c@{}}
\toprule
 & \textbf{Strong} & \textbf{Empirical} & \textbf{Batched critic} \\
\midrule
\textbf{RDP} 
  & {\cite{2011LiEtAlMainPaperOnDistributionPreservingQuantization,Dec2022WeakAndStrongPerceptionConstraintsAndRandomness}} 
  & {\cite{Sep2015RDPLimitedCommonRandomnessSaldi,Dec2022WeakAndStrongPerceptionConstraintsAndRandomness}} 
  & {\cite{2025OurISITRDPAlgorithmicRealism}} \\[3pt]
\textbf{Coordination} 
  & {\cite{2013PaulCuffDistributedChannelSynthesis}} 
  & {\cite{Cuff2010CoordinationCapacity}} 
  & \textbf{Open} \\
\bottomrule
\end{tabular}
\caption{The parallel between realism constraints (RDP) and 
coordination constraints across formulation types. References
indicate where the corresponding trade-off has been characterized in the point-to-point setting. 
The systematic correspondence across rows is a central theme of 
this paper; the open entry illustrates how advances in one row 
can suggest new problems in the other.}
\label{tab:constraint_taxonomy}
\end{table}

\subsection{Formal definitions of trade-offs}\label{subsec:unifying_definition_of_achievability}

The preceding discussion suggests that a ``good'' code may need to satisfy multiple, potentially competing objectives: fidelity to the source, realism of the reconstruction, or coordination between actions. We formalize these objectives through mathematical properties of the induced distribution $P^{(n)}$.

{Each of the constraints presented in Section \ref{subsec:constraints_toolbox} takes one of two forms:}
\begin{enumerate}
    \item A \emph{parametric constraint} $\mathcal{P}(n, P^{(n)}, \lambda)$ (e.g., \eqref{eq:def_distortion_n_letter} or \eqref{eq:def_strong_realism}) that must be satisfied for each sufficiently large $n$, where $\lambda$ is a parameter (e.g., a distortion level);
    \item An \emph{asymptotic property} $\mathcal{P}(\{P^{(n)}\}_{n \in \mathbb{N}})$ that pertains to the limiting behavior of the sequence of induced distributions as $n \to \infty$ (e.g., a convergence in probability such as \eqref{eq:typicality_convergence} or \eqref{eq:def_empirical_coordination}).

\end{enumerate}

The fundamental limits of compression correspond to the asymptotic regime where the blocklength $n$ tends to infinity. This leads to the following definition.

\begin{definition}[Achievability]\label{def:achievability_unified}
Consider a
set of constraints.
A tuple $(R, R_c, \lambda_1, \ldots, \lambda_K)$ is \emph{asymptotically achievable} if, for any $\varepsilon > 0$, there exists a sequence of $(n, R + \varepsilon, R_c + \varepsilon)$ codes whose induced distributions $\{P^{(n)}\}_n$ satisfy the constraints.
The tuple is \emph{achievable with deterministic schemes} if the codes can be chosen to be deterministic.
\end{definition}

This abstract formulation may seem overly general, but, as shown next, it provides a unified language for discussing rate-distortion theory, RDP trade-offs, and coordination problems.

\subsection{Classical Results as Special Cases}\label{subsec:clarifying_the_abstract_formulation}

To ground this abstract framework, we note how classical 
results emerge as special cases.\\
\textbf{Rate-distortion theory} uses one parametric 
constraint~\eqref{eq:def_distortion_n_letter} with 
additive distortion.
A pair $(R,\lambda)$ is achievable with deterministic schemes iff $R$ is at least:
\begin{IEEEeqnarray}{c}
R(\lambda) = \min_{p_{Y|X}: \mathbb{E}[d(X,Y)] 
\leq \lambda} I_p(X; Y)
\footnote{The mutual information between $X$ and $Y$ is a function of the joint distribution, here $p_{X}\cdot p_{Y|X},$ which appears as subscript in $I_p(X;Y).$}.
\end{IEEEeqnarray}
Whether a tuple is achievable does not depend on CR.\\
\textbf{The RDP trade-off} combines 
distortion~\eqref{eq:def_distortion_n_letter} and strong 
realism~\eqref{eq:def_strong_realism}; it generally 
requires $R_c > 0$.\\
\textbf{Channel synthesis} imposes 
constraint~\eqref{eq:def_strong_coordination} with 
$\lambda = 0$; the minimum rate is $R = I_q(X; Y)$, 
but achieving it requires CR at rate $R_c = H_q(Y|X)$.

\subsection{
Side Information and Remote Source 
Compression}

The basic setup extends naturally to scenarios with 
additional sources. We mention two important cases.

\begin{figure}[t!]
\begin{center}
\scalebox{0.69}{
\begin{tikzpicture}
  \node[box,minimum width=2.15cm] (Enc) at (0,0)    {Encoder};
  \node[box,minimum width=2.36cm] (Dec) at (5.05,0) {Decoder};

  \node[lbl] (P) at (-4.15,0.68) {$p_{X,Z}^{\otimes n}$};
  \node[lbl] (X) at (-2.13,1.22) {$X_{1:n}$};
  \node[lbl] (Z) at (-2.17,0)    {$Z_{1:n}$};
  \node[lbl] (Y) at (7.39,0)     {$Y_{1:n}$};
  \node[lbl] (M) at (2.48,0.37)  {$M\in[2^{nR}]$};
  \node[lbl] (J) at (2.48,-0.87) {$J\in[2^{nR_c}]$};

  \draw[wire]    (P.east)      -- (-3.01,0.68);
  \draw[wire]    (-3.01,0)     -- (-3.01,1.22);
  \draw[wire,ar] (-3.01,1.22)  -- (X.west);
  \draw[wire,ar] (-3.01,0)     -- (Z.west);
  \draw[wire,ar] (Z.east)      -- (Enc.west);

  \draw[wire,ar] (Enc.east)    -- (Dec.west);

  \draw[wire,ar] (J.west) -| (Enc.south);
  \draw[wire,ar] (J.east) -| (Dec.south);

  \draw[wire,ar] (Dec.east)    -- (Y.west);
\end{tikzpicture}
}
\end{center}
\caption{The remote source compression setting.}
\label{fig:remote_source_compression_n_letter}
\end{figure}
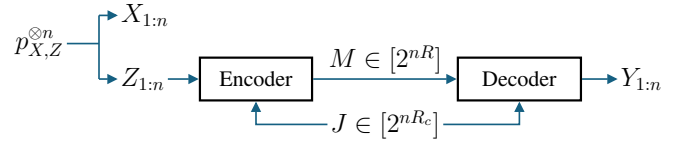

\paragraph{Remote Source Compression}
In some applications, the encoder does not observe the 
source $X_{1:n}$ directly, but only a noisy version 
$Z_{1:n}$ with $(X_{1:n}, Z_{1:n}) \sim p_{X,Z}^{\otimes n}$ 
(Figure~\ref{fig:remote_source_compression_n_letter}). 
The decoder must still produce a reconstruction $Y_{1:n}$ 
that is close to (and realistic with respect to) the 
\emph{remote} source $X_{1:n}$. This can be formalized 
similarly to Definition~\ref{def:n_letter_code}, and 
Definition~\ref{def:achievability_unified} applies. 
The constraints~\eqref{eq:def_distortion_n_letter} 
and~\eqref{eq:def_strong_realism} define the remote 
RDP trade-off.

\begin{figure}[t!]
\begin{center}
\scalebox{0.69}{
\begin{tikzpicture}
  \node[box,minimum width=2.50cm] (Enc) at (0,0)    {Encoder};
  \node[box,minimum width=2.72cm] (Dec) at (5.74,0) {Decoder};

  \node[lbl] (X) at (-2.58,0)    {$X_{1:n}$};
  \node[lbl] (Y) at (8.44,0)     {$Y_{1:n}$};
  \node[lbl] (M) at (2.82,-0.53) {$M\in[2^{nR}]$};
  \node[lbl] (J) at (2.92,0.89)  {$J\in[2^{nR_c}]$};
  \node[lbl] (Z) at (5.74,-1.25) {$Z_{1:n}$};

  \draw[wire,ar] (X.east)   -- (Enc.west);
  \draw[wire,ar] (Enc.east) -- (Dec.west);
  \draw[wire,ar] (Dec.east) -- (Y.west);

  \draw[wire,ar] (J.west) -| (Enc.north);
  \draw[wire,ar] (J.east) -| (Dec.north);

  \draw[wire,ar]        (Z.north) -- (Dec.south);
  \draw[wire,ar,dashed] (Z.west)  -| (Enc.south);
\end{tikzpicture}
}
\end{center}
\caption{Compression in the presence of side information.}
\label{fig:with_side_info_n_letter}
\end{figure}
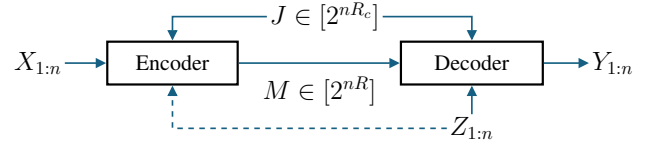

\paragraph{Side Information}
Suppose a second source $Z_{1:n}$, correlated with 
$X_{1:n}$ as $(X_{1:n}, Z_{1:n}) \sim p_{X,Z}^{\otimes n}$, 
is available at the decoder or at both terminals 
(Figure~\ref{fig:with_side_info_n_letter}). 
When combined with realism constraints, one may require
not just that $Y_{1:n}$ looks realistic, but that the 
pair $(Y_{1:n}, Z_{1:n})$ has the correct joint 
distribution:
\begin{IEEEeqnarray}{c}
\mathcal{P}_{\mathrm{joint}}(n, P^{(n)}, \lambda): 
\mathcal{D}_n\bigl(P^{(n)}_{Y_{1:n}, Z_{1:n}}, 
p_{X,Z}^{\otimes n}\bigr) \leq \lambda.
\label{eq:def_joint_realism}
\end{IEEEeqnarray}

The striking parallel between RDP and channel synthesis 
is the subject of the next section.

\section{Common characteristics of problems with distribution matching constraints}\label{sec:distribution_matching_everything}

In this section, we
present the known achievable regions for
RDP with strong realism constraints and channel synthesis
side by 
side.
Our goal is to build the reader's intuition that the achievable regions for these problems differ \emph{only} by the substitution $Y \to (X,Y),$ and that this pattern persists across multiple problem variants (point-to-point, remote source, side information).
For the information theorist, internalizing this pattern 
has immediate value: it provides a template for 
conjecturing achievable regions in new settings. 
Indeed, the productive connections documented in this 
paper were discovered precisely through this pattern: 
the characterization of the role of CR in the point-to-point 
RDP problem~\cite{Sep2015RDPLimitedCommonRandomnessSaldi} 
followed the proof structure of channel 
synthesis~\cite{2013PaulCuffDistributedChannelSynthesis},
and the same holds for the RDP problem in the presence of side information~\cite{OurJournalVersionRDPWithSideInfo}, where
the joint realism 
constraint~\eqref{eq:def_joint_realism} is also inspired by the strong coordination constraint~\eqref{eq:def_strong_coordination}.
We begin with the basic point-to-point setting before extending to scenarios with additional sources.

\subsection{The point-to-point scenario}\label{subsec:point_to_point_deep_similarities}

We consider
two problems formulated in Section~\ref{sec:unified_problem_formulation}: the RDP problem with a strong realism constraint, and channel synthesis. The former combines the distortion constraint~\eqref{eq:def_distortion_n_letter} with the strong realism constraint~\eqref{eq:def_strong_realism}; the latter imposes only the strong coordination constraint~\eqref{eq:def_strong_coordination}. Throughout this section, we assume the distortion measure is additive as in \eqref{eq:def_additive_distortion_function}.

\subsubsection{Why Randomization is Necessary}

A sharp departure from classical rate-distortion theory is that \emph{deterministic codes cannot satisfy stringent distribution matching constraints} {\cite{2011LiEtAlMainPaperOnDistributionPreservingQuantization,Dec2022WeakAndStrongPerceptionConstraintsAndRandomness}}.

For a deterministic code
at rate
$R < H(X)$ (the regime of interest for lossy compression),
we have
\begin{IEEEeqnarray}{c}
H_{P^{(n)}}(Y_{1:n}) \leq H_{P^{(n)}}(M) \leq nR < nH(X) = H(p_X^{\otimes n}).\nonumber
\end{IEEEeqnarray}
Thus, a deterministic code cannot satisfy
a strong realism constraint with $\lambda = 0,$
which requires $P^{(n)}_{Y_{1:n}} = p_X^{\otimes n}.$

The same argument applies to channel synthesis: if the target distribution $q_{Y|X}$ has $H_q(Y) > R$, then deterministic codes cannot achieve $\lambda = 0$ in~\eqref{eq:def_strong_coordination}.

To satisfy these constraints, the decoder must ``inject'' additional randomness to increase the entropy of its output. This randomness can come from two sources: local randomness at the decoder (which suffices for other problems with weaker realism constraints) or CR shared with the encoder (which is necessary for optimal performance). The amount of CR required is quantified in the achievable regions below.

\subsubsection{Achievable Region - The RDP Problem}

Consider the RDP problem with perfect strong realism ($\lambda_2 = 0$ in~\eqref{eq:def_strong_realism}). What tuples $(R, R_c, \lambda_1)$ of compression rate, CR rate, and distortion are achievable? The answer, established in~{\cite{Sep2015RDPLimitedCommonRandomnessSaldi}}, is that the closure of achievable tuples is the region:
\begin{align}
      & 
    \left\{ \begin{array}{rcl}
        \scalebox{1.0}{$\exists \  p_{Y,V|X}$} &:& \scalebox{1.0}{$p_{Y} = p_X$}\\
        \scalebox{1.0}{$R$} &\geq& \scalebox{1.0}{$I_p(X;V)$} \\
        \scalebox{1.0}{$R+R_c$} &\geq& \scalebox{1.0}{$I_p(Y;V)$} \\
        \scalebox{1.0}{$\lambda_1$} &\geq& \scalebox{1.0}{$\mathbb{E}_p[d(X, Y)]$}\\
        X \quad - &V& - \quad Y
    \end{array}\right\},
\label{region_RDP_near_perfect_strong_realism}
\end{align}
where $I_p$ denotes the mutual information according to the joint distribution $p,$ and $X-V-Y$ denotes that $(X,V,Y)$ form a Markov chain in that order.
This characterization holds for general alphabets $\mathcal{X}$ (e.g., $\mathbb{R}$) under mild assumptions {\cite{OurJournalVersionRDPWithSideInfo}}.
Hereafter, we only consider finite alphabets.

\subsubsection{Achievable Region - Channel Synthesis}

Now consider the problem of channel synthesis with perfect coordination  ($\lambda_1=0$ in~\eqref{eq:def_strong_coordination}). 
The closure of the set of pairs $(R,R_c),$ such that
$(R,R_c,\lambda_1=0)$ is asymptotically achievable, is {\cite{2013PaulCuffDistributedChannelSynthesis}}:
\begin{align}
      & 
    \left\{ \begin{array}{rcl}
        \scalebox{1.0}{$\exists \  p_{Y,V|X}$} &:& \scalebox{1.0}{$p_{X,Y} = q_{X,Y}$}\\
        \scalebox{1.0}{$R$} &\geq& \scalebox{1.0}{$I_p(X;V)$} \\
        \scalebox{1.0}{$R+R_c$} &\geq& \scalebox{1.0}{$I_p(X,Y;V)$} \\
        X \quad - &V& - \quad Y
    \end{array}\right\}.
\label{region_channel_synthesis_near_perfect}
\end{align}

{Comparing \eqref{region_RDP_near_perfect_strong_realism} and \eqref{region_channel_synthesis_near_perfect}, the only differences (besides the distortion constraint) are the distribution-matching equality and the bound on $R+R_c$: $Y$ is replaced by $(X,Y)$.} This is precisely because channel synthesis requires coordinating the joint distribution of input and output, while RDP with strong realism only requires the output marginal to match.

\subsubsection{Minimum Rates and the Role of Common Randomness}

What is the minimum compression rate for each problem, and how much CR is needed to achieve it?

For the RDP problem with distortion constraint $\lambda_1$, the minimum rate is~{\cite{2011LiEtAlMainPaperOnDistributionPreservingQuantization,Sep2015RDPLimitedCommonRandomnessSaldi}}:
\begin{IEEEeqnarray}{rCl}
R(\lambda_1) &=& \min_{\substack{p_{Y|X}: \, p_Y = p_X \\ \mathbb{E}_p[d(X,Y)] \leq \lambda_1}} I_p(X;Y).
\label{eq:RDP_function_perfect_realism}
\end{IEEEeqnarray}
This is analogous to Shannon's rate-distortion function, but with the additional constraint that the reconstruction distribution must match the source distribution.

For channel synthesis with target channel $q_{Y|X}$, the minimum rate is simply~\cite{2002WinterChannelSimulationUnlimitedCR
, 2013PaulCuffDistributedChannelSynthesis
}:
\begin{IEEEeqnarray}{rCl}
R(q_{Y|X}) &=& I_q(X;Y).
\label{eq:optimal_channel_synthesis_rate_is_MI}
\end{IEEEeqnarray}

Moreover, the minimum CR rate $R_c(\lambda_1)$ necessary to achieve the minimum compression rate $R(\lambda_1),$ and the minimum $R_c(q_{Y|X})$ necessary to achieve $R(q_{Y|X})$ have similar expressions
\cite[Proposition~1]{Sep2015RDPLimitedCommonRandomnessSaldi}:
\begin{IEEEeqnarray}{rCl}
R_c(\lambda_1) &=& \min_{\substack{p_{Y|X}
\text{ s.t. }
p_Y=p_X,
\ \mathbb{E}_p[d(X,Y)]\leq \lambda_1\\I_p(X;Y)=R(\lambda_1)}} H_p(Y{\dagger}X),
\IEEEeqnarraynumspace
\\
R_c(q_{Y|X}) &=& H_q(Y{\dagger}X),
\IEEEeqnarraynumspace
\end{IEEEeqnarray}
where $H(Y{\dagger}X)$ denotes the \textit{necessary conditional entropy} {\cite{Cuff2010CoordinationCapacity}}. Intuitively, this quantity measures the minimum amount of randomness the decoder needs, beyond what can be extracted from the message, to generate $Y$ with the correct conditional distribution given $X$. For many distributions of practical interest, $H(Y \dagger X) = H(Y|X)$, the standard conditional entropy {\cite{Cuff2010CoordinationCapacity,2013PaulCuffDistributedChannelSynthesis}}.
If $H(Y{\dagger}X)$ is replaced by $H(Y|X)$
above, we obtain
\begin{IEEEeqnarray}{rCl}
R_c(\lambda_1) &=& H_p(X) - R(\lambda_1), \label{eq:CR_rate_RDP} \\
R_c(q_{Y|X}) &=& H_q(Y) - R(q_{Y|X}). \label{eq:CR_rate_channel_synth}
\end{IEEEeqnarray}
where the equality $H_p(Y) = H_p(X)$ in~\eqref{eq:CR_rate_RDP} follows from the realism constraint $p_Y = p_X$.

\subsubsection{Practical Implications: CR Can Dominate}

Since in lossy compression the regime of interest is 
$R \ll H(X),$
equations~\eqref{eq:CR_rate_RDP} 
and~\eqref{eq:CR_rate_channel_synth} imply that:
\begin{IEEEeqnarray}{c}
R_c \gg R,
\end{IEEEeqnarray}
i.e., the required CR rate can far 
exceed the compression rate. {Concretely, the shared random seed must be much longer than the compressed representation, and any shortfall in CR must be compensated by increased communication rate, since $R + R_c \geq I(Y;V)$ must hold.}

This theoretical prediction, that strong realism requires substantial CR, stands in tension with practical experience: to the best of our knowledge, no existing neural codec has reported needing large amounts of shared randomness. {We return to this puzzle in Section~\ref{sec:algorithmic_realism}, where the framework of algorithmic realism offers a natural resolution.}

\medskip
\noindent\textbf{Remark.}
The CR requirement above is specific to
perfect strong realism ($\lambda_2 = 0$);
for
$\lambda_2 > 0$, 
substantial CR is 
conjectured~\cite{Dec2022WeakAndStrongPerceptionConstraintsAndRandomness} 
to remain necessary for optimal performance.
Under per-symbol 
realism~\eqref{eq:def_per_symbol_realism}, CR is not 
needed~\cite{Dec2022WeakAndStrongPerceptionConstraintsAndRandomness}; 
deterministic schemes suffice when $\lambda_2 > 0$, while 
decoder randomization is needed when $\lambda_2 = 0$. 
Under empirical realism~\eqref{eq:def_empirical_realism}, 
deterministic schemes also suffice.
{As mentioned in Section \ref{sec:unified_problem_formulation}, the tractability of the i.i.d.\ source model allows one to uncover properties of compression with realism constraints that we expect to raise relevant questions for practical scenarios.}
Such a conjecture is arguably less principled for per-symbol and empirical constraints
as they
are insensitive 
to the \emph{order} of symbols: producing only the two periodic sequences $0101\cdots$ and $1010\cdots$ with equal probability
would be deemed realistic for a Bernoulli($1/2$) 
source~{\cite{2025OurISITRDPAlgorithmicRealism}}.

\subsubsection{Common Analytical Tools}

The proofs of~\eqref{region_RDP_near_perfect_strong_realism} 
and~\eqref{region_channel_synthesis_near_perfect} 
\cite{Sep2015RDPLimitedCommonRandomnessSaldi,
2013PaulCuffDistributedChannelSynthesis} both rely on the 
\textit{soft covering lemma}, which ensures that random 
codebook selection induces approximately the correct output 
distribution (in contrast to the standard covering lemma~{\cite[Section~3.7]{2011BookElGamalKimNetworkInformationTheory}},
which ensures proximity to typical source sequences). 
Both schemes employ the \textit{likelihood 
encoder}~\cite{2016SongCuffLikelihoodEncoder}, which selects 
codewords with probabilities proportional to their likelihood 
under the source.
With sufficient CR rate, optimal RDP performance can be achieved by running an optimal channel 
synthesis scheme for the channel $p_{Y|X}$ attaining the 
minimum 
in~\eqref{eq:RDP_function_perfect_realism}~\cite{Sep2015RDPLimitedCommonRandomnessSaldi,TheisWagner2021VariableRateRDP,2022TheisGaussianDiffusion}.

\subsection{Scenarios with more sources and/or 
terminals}\label{subsec:beyond_the_basic_setup}

The parallels documented in 
Section~\ref{subsec:point_to_point_deep_similarities} 
extend to the settings with additional sources introduced 
in Section~\ref{sec:unified_problem_formulation}: remote 
source compression 
(Figure~\ref{fig:remote_source_compression_n_letter}) 
and compression with side information 
(Figure~\ref{fig:with_side_info_n_letter}).

\medskip\noindent\textbf{Remote source compression.}
{For perfect strong realism ($\lambda_2 = 0$), the closure of the set of achievable rates is given by~\cite[Theorem~1]{Sep2015RDPLimitedCommonRandomnessSaldi}:}
\begin{align}
      & 
    \left\{ \begin{array}{rcl}
        \scalebox{1.0}{$\exists \  p_{Y,V|Z,X}$} &:& \scalebox{1.0}{$p_{Y} = p_X$}\\
        \scalebox{1.0}{$R$} &\geq& \scalebox{1.0}{$I_p(Z;V)$} \\
        \scalebox{1.0}{$R+R_c$} &\geq& \scalebox{1.0}{$I_p(Y;V)$} \\
        \scalebox{1.0}{$\lambda_1$} &\geq& \scalebox{1.0}{$\mathbb{E}_p[d(X, Y)]$}\\
        X \ - \ Z &-& V \ - \ Y
    \end{array}\right\}.
\label{region_remote_RDP_near_perfect_strong_realism}
\end{align}
Similarly, the constraint \eqref{eq:def_strong_coordination} defines a trade-off between $R,$ $R_c,$ and $\lambda_1$ called the remote channel synthesis trade-off \cite{2025OurITWRemoteChannelSynthesis}.
{When $\lambda_1=0,$ the closure of the set of achievable tuples is the following region \cite{2025OurITWRemoteChannelSynthesis}:}
\begin{align}
      & 
    \left\{ \begin{array}{rcl}
        \scalebox{1.0}{$\exists \  p_{Y,V|Z,X}$} &:& \scalebox{1.0}{$p_{X,Y} = q_{X,Y}$}\\
        \scalebox{1.0}{$R$} &\geq& \scalebox{1.0}{$I_p(Z;V)$} \\
        \scalebox{1.0}{$R+R_c$} &\geq& \scalebox{1.0}{$I_p(X,Y;V)$} \\
        X \ - \ Z &-& V \ - \ Y
    \end{array}\right\}.
\label{region_remote_channel_synthesis_near_perfect}
\end{align}
{Thus, the optimal trade-offs in the remote source setting are likewise closely related.} {Moreover, it can be shown that when the CR rate is sufficiently large, the expressions for the optimal rate resemble \eqref{eq:RDP_function_perfect_realism} and \eqref{eq:optimal_channel_synthesis_rate_is_MI}.}

\medskip\noindent\textbf{Side information.}
We present the case where $Z_{1:n}$ is available at 
both terminals. Constraints~\eqref{eq:def_distortion_n_letter} 
and~\eqref{eq:def_joint_realism} define a trade-off
characterized by
the following region when $\lambda_2 = 0$ 
\cite{OurJournalVersionRDPWithSideInfo}:
\begin{align}
      & 
    \left\{ \begin{array}{rcl}
        \scalebox{1.0}{$\exists \  p_{Y,V|Z,X}$} &:& \scalebox{1.0}{$p_{Y,Z} = p_{X,Z}$}\\
        \scalebox{1.0}{$R$} &\geq& \scalebox{1.0}{$I_p(X;V|Z)$} \\
        \scalebox{1.0}{$R+R_c$} &\geq& \scalebox{1.0}{$I_p(Y;V|Z)$} \\
        \scalebox{1.0}{$\lambda_1$} &\geq& \scalebox{1.0}{$\mathbb{E}_p[d(X, Y)]$}\\
        X \ \ - & (Z , V) & - \ \ Y
    \end{array}\right\}.
\label{region_RDP_near_perfect_strong_joint_realism_with_SI_two_sided}
\end{align}

Similarly, the following
constraint defines a trade-off:
\begin{IEEEeqnarray}{c}
\mathcal{P}_1(n,P^{(n)},\lambda_1): \
\mathcal{D}_n(P^{(n)}_{X_{1:n},Y_{1:n},Z_{1:n}},q_{X,Y,Z}^{\otimes n})
\leq \lambda_1
.
\label{eq:def_joint_strong_coordination_with_SI}
\IEEEeqnarraynumspace
\end{IEEEeqnarray}
When
$\lambda_1=0,$
the
closure of the set of
achievable

rates
is \cite{2015TITYassaeeEtAlChannelSimulationInteractiveComm}:
\begin{align}
      & 
    \left\{ \begin{array}{rcl}
        \scalebox{1.0}{$\exists \  p_{Y,V|Z,X}$} &:& \scalebox{1.0}{$p_{X,Y,Z} = q_{X,Y,Z}$}\\
        \scalebox{1.0}{$R$} &\geq& \scalebox{1.0}{$I_p(X;V|Z)$} \\
        \scalebox{1.0}{$R+R_c$} &\geq& \scalebox{1.0}{$I_p(X,Y;V|Z)$} \\
        X \ \ - & (Z , V) & - \ \ Y
    \end{array}\right\}.
\label{region_near_perfect_strong_joint_coordination_with_SI_two_sided}
\end{align}
Therefore, once again, the optimal trade-offs for the RDP problem and the distributed coordination problem are closely analogous. When $Z_{1:n}$ is only available at the decoder, the above characterizations of the achievable tuples still hold provided that the additional requirement $Z-X-V$ is added to each of \eqref{region_RDP_near_perfect_strong_joint_realism_with_SI_two_sided} and \eqref{region_near_perfect_strong_joint_coordination_with_SI_two_sided} \cite{2015TITYassaeeEtAlChannelSimulationInteractiveComm,OurJournalVersionRDPWithSideInfo},
and the same close correspondence continues to hold.

\subsubsection{Open Problems, Extensions}

The tight connections described above suggest fertile 
ground for research.\\
\textbf{Connections to optimal transport.}
The optimal transport (OT) problem seeks, given two 
distributions $\mu$ and $\nu$ and a cost function $c$, 
the coupling with marginals $\mu$ and $\nu$ that 
minimizes the expected 
cost~\cite{Villani2009OptimalTransport}. 
The RDP function~\eqref{eq:RDP_function_perfect_realism} 
has a related structure: it optimizes over couplings of 
$p_X$ with itself under a distortion budget 
$\mathbb{E}[d(X,Y)] \leq \lambda_1$. 
{When the target reconstruction distribution differs from the source, the connection becomes even more explicit, and, together with its coordination counterpart --- remote channel synthesis~{\cite{2025OurITWRemoteChannelSynthesis}} --- it provides a framework for studying generalized transport under communication constraints~{\cite{2022RDPasOptimalTransport}}.}\\
\textbf{Multi-terminal settings.}
Do similar parallels hold for broadcast and
multiple access channels, or networks with more than 
two agents? Initial work appears 
in~{\cite{2024OurNonInteractiveCoordinationNeurIPSWorkshop}}.\\
\textbf{Successive refinement and multiple descriptions.}
RDP problems in these settings have been studied, 
but connections to coordination theory remain unexplored.\\
\textbf{Partial coordination.}
The case where side information is present and only 
$(X_{1:n}, Y_{1:n})$ (not $Z_{1:n}$) must be coordinated 
remains 
open~{\cite{2020ISITStrongCoordinationWithCodedSI}}.

In the following section, we turn to recent developments that address limitations of the distribution matching framework
as a model of realism.

\section{Batched critics and algorithmic realism}
\label{sec:algorithmic_realism}

The preceding section focused on distributional constraints, 
{which have a conceptual limitation:} under realism
constraints 
like~\eqref{eq:def_strong_realism} 
and~\eqref{eq:def_per_symbol_realism}, it is not meaningful 
to speak of whether an individual realization is 
realistic~{\cite{2025OurISITRDPAlgorithmicRealism}}; 
realism is a property of the ensemble. 
Yet human observers judge realism one image at a time. 
This motivates the use of a \emph{realization-based} metric 
(critic) to formalize realism:
$Y_{1:n} \mapsto \delta(Y_{1:n}) \in\mathbb{R}.$ One may also consider a batched critic, i.e., a function that jointly inspects multiple samples: $
(
Y^{(1)}_{1{:}n},
\ldots,
Y^{(B_n)}_{1{:}n}
) \mapsto\delta
(
Y^{(1)}_{1{:}n},
\ldots,
Y^{(B_n)}_{1{:}n}
)$
(cf.\ Section~\ref{sec:unified_problem_formulation}).
In this section, we discuss the use of batched critics to model both realism and coordination.
We start by providing examples of realization-based metrics for realism.

For an i.i.d.\ source, what properties should a good 
critic detect in reconstructions? Consider a
$\text{Bernoulli}(1/2)$ source. A critic might measure 
how much the frequency of ones departs from $1/2$, relative 
to the typical fluctuation $\sqrt{n}$. For true source 
sequences, this ratio stays bounded; but if a coding scheme 
produces reconstructions whose frequency of ones departs 
from $1/2$ by more than $n^{-(1/2-\epsilon)}$ for some 
$\epsilon > 0$, the critic's expected output grows without 
bound as $n$ 
increases.
Similarly, a critic could inspect the frequency of the block $01,$ or any other pattern.
Note that such critics only test the empirical distribution 
and are insensitive to the ordering of symbols. A more 
powerful critic can inspect the length of the longest run 
of consecutive ones: for an i.i.d.\ Bernoulli($1/2$) 
sequence of length $n$, this length concentrates around 
$\log_2 n$. If a coding scheme produces reconstructions 
whose longest run grows at a different rate,
the expected output 
of this critic diverges with $n$. More generally, any limit 
theorem satisfied by the source distribution can be 
embodied in a critic that diverges when the theorem is 
violated~{\cite{2025OurISITRDPAlgorithmicRealism}}. In other 
words, satisfying a given limit theorem can be formalized as 
$\sup_n \mathbb{E}[\delta(\cdot)] < \infty$ for some 
critic $\delta.$

Each RDP (resp. coordination) problem mentioned in previous sections involved only one realism (resp. coordination) constraint. {However, one can impose multiple constraints, e.g., by selecting some of the examples of critics described above and imposing a constraint for each critic; while each of these examples arguably does not suffice to model realism, combining constraints from a sufficiently rich family may yield a strong model.}
Formally, consider $B_n \in\mathbb{N},$ and let $\mathcal{C}$ denote a family of batched critics with batch size $B_n.$
Given a constraint for each critic, and a potential distortion constraint, one obtains a trade-off (Definition \ref{def:achievability_unified}).
For $B_n = 1$, the critics 
inspect individual samples; as $B_n$ grows, they 
can detect increasingly subtle distributional properties.
The role of randomness (including CR) may vary dramatically across these different cases.
{In Subsection \ref{subsec:alg_realism}, we provide an example of a class $\mathcal{C}$ of critics and a corresponding RDP problem for which such a phenomenon holds {\cite{2025OurISITRDPAlgorithmicRealism}}.}
In Subsection~\ref{subsec:detailed_proposal_open_problem}, we propose to apply this paradigm to coordination, once again transferring a development across the parallel.

\subsection{Algorithmic realism}\label{subsec:alg_realism}

A particularly powerful choice of $\mathcal{C}$ is provided by algorithmic information theory~\cite{2024TheisUniversalCriticsPositionPaper,BookKolmogorovComplexity}. One can think 
of this class as a vast set that includes tests for all 
limit theorems.
See~{\cite{2025OurISITRDPAlgorithmicRealism}} for a comprehensive exposition, and~{\cite{2024TheisUniversalCriticsPositionPaper}} for a lucid high-level presentation. 
In~{\cite{2024TheisUniversalCriticsPositionPaper}}, the 
author argues in favor of measuring realism via a batched versions of these critics
in a variety of problems, such as 
outlier detection and generative modeling.

In~{\cite{2025OurISITRDPAlgorithmicRealism}}, the RDP 
problem is studied with finite alphabets, a distortion 
constraint~\eqref{eq:def_distortion_n_letter} with an 
additive distortion function, and the following realism 
constraint, given a sequence $\{B_n\}_{n\in\mathbb{N}}$ 
of batch sizes: for every critic 
$\delta \in \mathcal{C}$,
\begin{IEEEeqnarray}{c}
\sup_{n\in\mathbb{N}}
\mathbb{E}
\big[
\delta
(
Y^{(1)}_{1{:}n},
\ldots,
Y^{(B_n)}_{1{:}n}
)
\big]
<\infty,
\label{eq:def_batched_critic_constraint}
\end{IEEEeqnarray}
where $\{Y^{(k)}_{1{:}n}\}_{k \in [B_n]}$ is a batch 
of i.i.d.\ samples from $P^{(n)}_{Y_{1{:}n}}.$ 
This formally defines a trade-off between $R, R_c, 
\lambda_1$ (Definition~\ref{def:achievability_unified}). 
Satisfying~\eqref{eq:def_batched_critic_constraint} 
requires the reconstructions to be highly realistic: 
they must asymptotically satisfy a vast range of 
algorithmically testable properties of 
$\{p_X^{\otimes n}\}_{n\in\mathbb{N}},$ including the 
examples discussed above --- which were also of the form $\sup_n \mathbb{E}[\delta(\cdot)] < \infty$ --- and essentially all limit 
theorems.

The main result 
in~{\cite{2025OurISITRDPAlgorithmicRealism}} is that if 
$B_n$ does not grow exponentially fast with $n,$ then 
deterministic schemes can achieve the optimal trade-off. 
Moreover, if $B_n$ grows fast enough, the optimal 
trade-off is identical to the one presented in 
Section~\ref{sec:distribution_matching_everything}, and 
the region is given 
by~\eqref{region_RDP_near_perfect_strong_realism}. 
Intuitively, jointly inspecting a very large number of 
reconstructions is equivalent to inspecting the 
reconstruction distribution $P^{(n)}_{Y_{1:n}}.$ 
This provides a compelling explanation for the fact that 
no need for large amounts of CR has been observed in 
state-of-the-art codecs.

\subsection{An open problem: batched critics for 
coordination}\label{subsec:detailed_proposal_open_problem}

The framework above suggests an analogous formulation 
for coordination: for certain applications, the coordination requirement may be more suitably formalized by batched critics, and this may result in a fundamentally different trade-off compared to standard formulations. Consider the channel synthesis setup 
of Section~\ref{sec:distribution_matching_everything},
but replace the strong coordination 
constraint~\eqref{eq:def_strong_coordination} with a 
realization-based requirement of the following form: for a distribution $P^{(n)}$ induced by a scheme, consider a constraint for each critic in a class $\mathcal{C}$ 
of batched critics, each taking as input a batch of $B_n$ i.i.d. pairs $(X_{1:n}^{(k)},Y_{1:n}^{(k)})$
sampled from $P^{(n)}.$
The resulting trade-off is expected to inherit the interpolation property described above: for $B_n = 1$, deterministic codes suffice, while for large enough $B_n$, strong coordination is recovered and CR becomes necessary. Characterizing the 
role of randomness as a function of $B_n$ constitutes an open problem 
(Table~\ref{tab:constraint_taxonomy}).

This formulation is the $Y \to (X,Y)$ translation of 
the RDP formulation described above (see also~{\cite{2025OurISITRDPAlgorithmicRealism}}), but
its analysis may involve additional technical difficulties since $X$ is not under 
the designer's control.
If the class $\mathcal{C}$ is rich enough, the coordination constraint would be much stronger than empirical 
coordination~\eqref{eq:def_empirical_coordination}, 
which only pertains to frequency statistics. {A canonical choice of $\mathcal{C}$ is the class of randomness tests from algorithmic information theory introduced above: requiring that realizations pass all algorithmic tests for being i.i.d.\ according to the target $q_{X,Y}$ provides a natural realization-based interpolation of the notion of strong coordination.}

\section{Conclusion and Future Directions}

We have shown that lossy compression with 
realism constraints and distributed coordination are 
connected in ways that are not merely aesthetically 
appealing but concretely productive. 
The RDP trade-off with strong realism constraints and channel synthesis have nearly identical information-theoretic characterizations.
These connections 
have guided the characterization of the role of CR in 
RDP problems and suggested new constraint definitions for 
compression with side information.
We have also surveyed recent developments, including batched critics and algorithmic realism, and shown how advances on the RDP problem naturally suggest new problems for coordination theory, and vice versa.

The connections revealed here open
several
avenues for future research.
The extension of batched critics to coordination 
(Table~\ref{tab:constraint_taxonomy}) is one concrete instance; 
more broadly, multi-terminal networks (broadcast, multiple access, 
relay), successive refinement, and connections to optimal transport 
theory all offer opportunities for the parallel to generate new 
results.
As AI moves toward systems of multiple interacting agents, 
the theory of distributed coordination provides a principled 
framework for understanding fundamental limits under 
communication constraints, and the bridge to compression 
theory documented here may accelerate progress on both sides.

\section*{Acknowledgment}
This work received funding from
EU Horizon 2020 MSC-ITN Greenedge (GA. No. 953775),
from UKRI for projects INFORMED-AI (EP/Y028732/1),  AI-R (ERC Consolidator Grant, EP/X030806/1) and NSF-EPSRC Neural JSCC (UKRI1463), and from
SNS JU for project 6G-GOALS (GA. No. 101139232).

\textbf{Yassine Hamdi}
(Member, IEEE)
is a Research Associate in Machine Learning and Wireless Communications
at Imperial College London,
working with Pr. Deniz G\"{u}nd\"{u}z (Imperial College) and Pr. Ioannis Kontoyiannis (Cambridge University)
as part of the UKRI project INFORMED-AI which focuses on the intersection between information theory and machine learning.
He was
a PhD student
at the Intelligent Systems and Networks Group
at Imperial College London
from 2021 to 2026.
He graduated from Ecole Polytechnique (Palaiseau, France) in 2020, where he undertook research in applied probability theory. He received his MSc degree from the University of Paris-Saclay (France) in Statistics and Machine Learning in 2021.
His research interests include
realism constraints in
lossy compression and the rate-distortion-perception trade-off, as well as coordination constraints in networks, and diffusion models theory.

\textbf{Deniz G\"{u}nd\"{u}z}
(Fellow, IEEE)
received
the
Ph.D.
degree in electrical engineering from the NYU
Tandon School of Engineering in
2007.
In 2012, he joined
Imperial College London,
U.K., where he is currently a Professor of information
processing.
He is also
an Amazon Scholar.
He held positions with
Princeton University as a Post-Doctoral Researcher
and Stanford University as a Research Assistant Professor from
2007 to 2009.
He is
an Elected Member
of the IEEE
SPCOM
and
MLSP
Technical Committees.
He received
the IEEE CommSoc
CTTC
Early Achievement Award in 2017,
and the
Starting (2016), Consolidator (2022),
and Proof-of-Concept (2023)
Grants
of European Research Council (ERC).
He served
in editorial roles for
IEEE TRANS. INF. THEORY,
IEEE TRANS. COMMUN., 
IEEE TRANS. WIRELESS COMMUN.,
IEEE J. SEL. AREAS COMMUN.,
and IEEE J. SEL. AREAS INF. THEORY.


\begin{thebibliography}{10}
\providecommand{\url}[1]{#1}
\csname url@samestyle\endcsname
\providecommand{\newblock}{\relax}
\providecommand{\bibinfo}[2]{#2}
\providecommand{\BIBentrySTDinterwordspacing}{\spaceskip=0pt\relax}
\providecommand{\BIBentryALTinterwordstretchfactor}{4}
\providecommand{\BIBentryALTinterwordspacing}{\spaceskip=\fontdimen2\font plus
\BIBentryALTinterwordstretchfactor\fontdimen3\font minus \fontdimen4\font\relax}
\providecommand{\BIBforeignlanguage}[2]{{%
\expandafter\ifx\csname l@#1\endcsname\relax
\typeout{** WARNING: IEEEtran.bst: No hyphenation pattern has been}%
\typeout{** loaded for the language `#1'. Using the pattern for}%
\typeout{** the default language instead.}%
\else
\language=\csname l@#1\endcsname
\fi
#2}}
\providecommand{\BIBdecl}{\relax}
\BIBdecl

\bibitem{2023GunduzSurveyContextAndSemanticsAndTaskOriented}
D.~Gündüz \emph{et~al.}, ``{Beyond Transmitting Bits: Context, Semantics, and Task-Oriented Communications},'' \emph{IEEE Journal on Selected Areas in Communications}, vol.~41, no.~1, 2023.

\bibitem{2011BookElGamalKimNetworkInformationTheory}
A.~El~Gamal and Y.~Kim, \emph{{Network Information Theory}}.\hskip 1em plus 0.5em minus 0.4em\relax {Cambridge (UK)}: {Cambridge University Press}, 2011.

\bibitem{2011LiEtAlMainPaperOnDistributionPreservingQuantization}
M.~Li \emph{et~al.}, ``{On Distribution Preserving Quantization},'' 2011, arXiv:1108.3728.

\bibitem{2019BlauMichaeliRethinkingLossyCompressionTheRDPTradeoff}
Y.~Blau and T.~Michaeli, ``{Rethinking Lossy Compression: The Rate-Distortion-Perception Tradeoff},'' in \emph{{ICML}}, 2019.

\bibitem{2024TheisUniversalCriticsPositionPaper}
L.~Theis, ``{Position: What Makes an Image Realistic?}'' in \emph{ICML}, 2024.

\bibitem{Cuff2010CoordinationCapacity}
P.~Cuff \emph{et~al.}, ``{Coordination Capacity},'' \emph{IEEE Transactions on Information Theory}, vol.~56, no.~9, 2010.

\bibitem{2013PaulCuffDistributedChannelSynthesis}
P.~Cuff, ``{Distributed Channel Synthesis},'' \emph{{IEEE Transactions on Information Theory}}, vol.~59, no.~11, 2013.

\bibitem{Sep2015RDPLimitedCommonRandomnessSaldi}
N.~Saldi \emph{et~al.}, ``{Output Constrained Lossy Source Coding With Limited Common Randomness},'' \emph{{IEEE Transactions on Information Theory}}, vol.~61, no.~9, 2015.

\bibitem{OurJournalVersionRDPWithSideInfo}
Y.~{Hamdi} \emph{et~al.}, ``{Rate-Distortion-Perception Trade-off with Strong Realism Constraints: Role of Side Information and Common Randomness},'' \emph{IEEE Transactions on Information Theory}, 2026, {Early Access, DOI: 10.1109/TIT.2026.3702694}.

\bibitem{2025OurISITRDPAlgorithmicRealism}
Y.~Hamdi \emph{et~al.}, ``{The Rate-Distortion-Perception Trade-Off with Algorithmic Realism},'' in \emph{IEEE International Symposium on Information Theory}, 2025.

\bibitem{2003SSIM}
Z.~Wang \emph{et~al.}, ``{Multiscale Structural Similarity for Image Quality Assessment},'' in \emph{Asilomar Conference on Signals, Systems \& Computers}, 2003.

\bibitem{2018LPIPS}
R.~Zhang \emph{et~al.}, ``{The Unreasonable Effectiveness of Deep Features as a Perceptual Metric},'' in \emph{IEEE/CVF CVPR Conference}, 2018.

\bibitem{2024ISITYangQiuAaronWagnerWassersteinDistortionFirstPublicationOfItsThoereticalProperties}
Y.~Qiu and A.~B. Wagner, ``{Low-Rate, Low-Distortion Compression with Wasserstein Distortion},'' in \emph{IEEE International Symposium on Information Theory}, 2024.

\bibitem{2025GoogleWassersteinDistortionToImproveOverfittedCompressor}
J.~Ball\'e \emph{et~al.}, ``{Good, Cheap, and Fast: Overfitted Image Compression with Wasserstein Distortion},'' in \emph{IEEE/CVF CVPR Conference}, 2025.

\bibitem{2022PoELIC}
D.~He \emph{et~al.}, ``{PO-ELIC: Perception-Oriented Efficient Learned Image Coding},'' in \emph{IEEE/CVF CVPR Workshops}, 2022.

\bibitem{2019AgustssonMentzerGANforExtremeCompression}
E.~Agustsson \emph{et~al.}, ``{Generative Adversarial Networks for Extreme Learned Image Compression},'' in \emph{{IEEE/CVF CVPR Conference}}, 2019.

\bibitem{2022TheisGaussianDiffusion}
L.~Theis \emph{et~al.}, ``{Lossy Compression with Gaussian Diffusion},'' \emph{{arXiv e-prints}}, 2022, arxiv:2206.08889.

\bibitem{Dec2022WeakAndStrongPerceptionConstraintsAndRandomness}
J.~Chen \emph{et~al.}, ``{On the Rate-Distortion-Perception Function},'' \emph{{IEEE Journal on Selected Areas in Information Theory}}, vol.~3, no.~4, 2022.

\bibitem{TheisWagner2021VariableRateRDP}
L.~Theis and A.~B. Wagner, ``{A Coding Theorem for the Rate-Distortion-Perception Function},'' in \emph{{Neural Compresison Workshop @ ICML}}, 2021.

\bibitem{1973GacsKornerCommonInformation}
P.~G{\'a}cs and J.~Korner, ``{Common Information Is Far Less Than Mutual Information},'' \emph{Problems of Control and Information Theory}, vol.~2, 1973.

\bibitem{1975WynerCommonInformationAndSoftCoveringLemma}
A.~Wyner, ``{The Common Information of Two Dependent Random Variables},'' \emph{IEEE Transactions on Information Theory}, vol.~21, no.~2, 1975.

\bibitem{2002WinterChannelSimulationUnlimitedCR}
A.~Winter, ``{Compression of Sources of Probability Distributions and Density Operators},'' 2002, arXiv:quant-ph/0208131.

\bibitem{2016SongCuffLikelihoodEncoder}
E.~C. Song \emph{et~al.}, ``{The Likelihood Encoder for Lossy Compression},'' \emph{{IEEE Transactions on Information Theory}}, vol.~62, no.~4, 2016.

\bibitem{2025OurITWRemoteChannelSynthesis}
Y.~Hamdi and D.~G\"{u}nd\"{u}z, ``{Remote Channel Synthesis},'' in \emph{IEEE Information Theory Workshop}, 2025.

\bibitem{2015TITYassaeeEtAlChannelSimulationInteractiveComm}
M.~H. Yassaee, A.~Gohari, and M.~R. Aref, ``{Channel Simulation via Interactive Communications},'' \emph{{IEEE Transactions on Information Theory}}, vol.~61, no.~6, 2015.

\bibitem{Villani2009OptimalTransport}
C.~Villani, \emph{{Optimal Transport: Old and New}}.\hskip 1em plus 0.5em minus 0.4em\relax Springer, 2009.

\bibitem{2022RDPasOptimalTransport}
H.~Liu \emph{et~al.}, ``{Cross-Domain Lossy Compression as Entropy Constrained Optimal Transport},'' \emph{IEEE Journal on Selected Areas in Information Theory}, 2022.

\bibitem{2024OurNonInteractiveCoordinationNeurIPSWorkshop}
Y.~Hamdi \emph{et~al.}, ``{Non-interactive Remote Coordination},'' in \emph{Workshop on Machine Learning and Compression @ NeurIPS}, 2025.

\bibitem{2020ISITStrongCoordinationWithCodedSI}
V.~Ramachandran \emph{et~al.}, ``{Strong Coordination with Side Information},'' in \emph{{IEEE International Symposium on Information Theory}}, 2020.

\bibitem{BookKolmogorovComplexity}
M.~Li and P.~Vitányi, \emph{{An Introduction to Kolmogorov Complexity and Its Applications}}, 4th~ed.\hskip 1em plus 0.5em minus 0.4em\relax Springer, 2019.

\end{thebibliography}
\end{document}